\documentclass[aps,twocolumn,amsmath,amssymb,preprintnumbers]{revtex4}
\usepackage{amsmath} \usepackage{amsfonts} \usepackage{amssymb}
\usepackage{titlesec}
\usepackage{bbm}
\usepackage{epsfig}
\usepackage{graphics}
\usepackage{graphicx}
\newcommand{\be}{\begin{equation}}
\newcommand{\ee}{\end{equation}}
\newcommand{\ba}{\begin{eqnarray}}
\newcommand{\ea}{\end{eqnarray}}
\newcommand{\nn}{\nonumber}
\newcommand{\kr}{\rangle}
\newcommand{\kl}{\langle}
\titleformat{\subsection}[block]{\normalfont\bfseries}{\thesubsection.}{1ex}{}
\titlespacing{\subsection}{0pt}{10pt}{1pt}[0pt]
\titleformat*{\section}{\large\bfseries}
\renewcommand{\thesubsection}{\arabic{subsection}}

\begin{document}

\title[ ]{Hot big bang or slow freeze?}

\author{C. Wetterich}
\affiliation{Institut  f\"ur Theoretische Physik\\
Universit\"at Heidelberg\\
Philosophenweg 16, D-69120 Heidelberg}

\begin{abstract}
We confront the big bang for the beginning of the universe with an equivalent picture of a slow freeze - a very cold and slowly evolving universe. In the freeze picture the masses of elementary particles increase and the gravitational constant decreases with cosmic time, while the Newtonian attraction remains unchanged. The freeze and big bang pictures both describe the same observations or physical reality. We present a simple ``crossover model'' without a big bang singularity. In the infinite past space-time is flat. Our model is compatible with present observations, describing the generation of primordial density fluctuations during inflation as well as the present transition to a dark energy dominated universe. 

\end{abstract}

\maketitle

The early stages in the evolution of our universe are generally depicted as a big bang. The temperature of an early plasma of radiation and particles was much higher than the temperature of 2.7K measured in the cosmic microwave background (CMB), exceeding in early stages by far the temperature of the sun or any other object in the present universe. This fireball resulted from a type of extremely fast primordial explosion - the big bang. Characteristic time scales of the early stages of the explosion were $10^{-30}$ seconds or shorter, extremely tiny as compared to the present time scale of the cosmic expansion of around $10^{10}$ years. 

In this note we contrast the big bang picture with a very different alternative picture of a slow freeze. We present a specific ``crossover model'' which is described both in the freeze and big bang pictures. In the freeze picture the characteristic mass scale is set by a parameter
\be\label{AA1}
\mu=2\cdot 10^{-33}{\rm eV}.
\ee
This parameter is about the value that the present Hubble parameter takes in the big bang picture. In contrast to the big bang picture, however, the time scale $\mu^{-1}=10^{10}$ yr $(\hbar=c=k_B=1)$ characterizes the evolution of the universe during the radiation- and matter dominated epochs as well. For the inflationary epoch in primordial cosmology the characteristic time scale increases to even larger values, tending to infinity in the infinite past. The evolution of the universe has always been very slow. The cosmological solution can be continued to the infinite past. No big bang singularity is present. 

In the freeze picture the universe is shrinking rather than expanding during the radiation- and matter dominated epochs \cite{CWU}. Correspondingly, the temperature decreases if we look back in time - the early universe was an extremely cold place. At the time when the CMB was emitted the temperature of the plasma was only $82$ mK. And the universe was even much colder further in the past. Looking backwards in time we may associate the early stages of the universe with a state of ``freeze'' from which the universe is very slowly thawing. 

Despite the striking differences to the big bang picture for the evolution of geometry and temperature, this freeze picture is compatible with all present cosmological and experimental observations. The crucial ingredient is the increase of all particle masses as well as the Planck mass, induced by a scalar field $\chi$ whose value increases monotonically. In the infinite past $\chi$ goes to zero, while its present value has reached the reduced Planck mass $\chi(t_0)=M=2.44\cdot 10^{18}$GeV. In our normalization $\chi$ can be associated directly with the variable Planck mass. The mass of the electron $m_e$ or the proton $m_p$ is proportional to $\chi$. The strict observational bounds on a time variation of the ratio between nucleon mass and Planck mass, or the ratio $m_e/m_p$, are obeyed. The electromagnetic fine structure constant does not depend on $\chi$, such that atomic binding energies scale $\sim m_e\sim \chi$. 

Looking towards the past the electron mass decreases even faster than the temperature. At the time of the CMB emission one has $m_e\approx 14.2$eV, such that $T/m_e=5\cdot 10^{-7}$, with 
the 
binding energy of hydrogen  around 50 times the temperature. The size of the hydrogen atom at this moment is $1.9\mu m$, a factor 36000 larger than the present Bohr radius. The scale factor at least scattering was larger than today, $a_{ls}/a_0\approx 33$. However, the ratio of the scale factor divided by the size of the hydrogen atom was a factor 1091 smaller than at present, the same as in the big bang picture. Physical observables are dimensionless and can therefore depend only on dimensionless ratios of masses or lengths. Thus the big bang and freeze pictures can describe the same physical reality. 

The potential and kinetic energy of the homogeneous scalar field $\chi(t)$ can be associated with dynamical dark energy \cite{CW3,RP}. The scalar field $\chi$ plays the role of the cosmon. Our model contains no fixed parameter for the gravitational constant. The Planck mass increases with time and is huge at present due to a long exponential increase of $\chi(t)$. The tiny ratio of the present dark energy density divided by the fourth power of the Planck mass, $\rho_h(t_0)/\chi^4(t_0)=\rho_h(t_0)/M^4$, is explained dynamically and does not require any tuning of parameters. In the early stages of cosmology the same scalar field $\chi$ acts as the inflaton. Our model realizes``cosmon inflation'' \cite{CI}. 

The proposed crossover model can be described equivalently in a big bang picture. This is achieved by a Weyl scaling \cite{Weyl,Di} of the metric. In the resulting ``Einstein frame'' the Planck mass or $m_e$ and $m_p$ do no longer depend on time. In this frame our model becomes a standard quintessence model with an exponential potential. Also inflation takes a familiar form. Physical observables do not depend on the choice of frame \cite{CW1,DamE,FR3,Cat1,DeS} (``field relativity'' \cite{CWU}). They are often computed most easily in the Einstein frame. The naturalness of our model is, however, better understood in the freeze frame. 

We emphasize that the ratio between temperature and the electron mass was higher in the past than today in both pictures. In this relative sense the ``hot plasma'' inferred from nucleosynthesis or the CMB is realized in nature, independently of the picture. When we compare the temperature of the plasma to the present temperature of the CMB the possible time 
evolution of the electron mass enters, however. This leads in the freeze picture to a plasma temperature much smaller than the present CMB-temperature.

In this note we investigate a very simple model which involves only three dimensionless parameters besides the masses and couplings of the particles of the standard model of particle physics. It is compatible with all present cosmological observations, ranging from primordial density fluctuations to the present properties of dark energy. Involving no more free parameters than the $\Lambda$CDM model of a cosmological constant, our model is subject to many observational tests and possible falsification.

{\em Crossover model.} The coupled cosmon-gravity system of our model is specified by the quantum effective action 
\be\label{1}
\Gamma=\int d^4x\sqrt{g}
\left\{-\frac{\chi^2}{2}R+
\left(\frac{2}{\alpha^2}-3\right)
\partial^\mu \chi\partial_\mu\chi+V(\chi)\right\},
\ee
from which the field equations for the metric and the cosmon follow by variation. The metric $g_{\mu\nu}$ appears in the curvature scalar $R,~\partial^\mu=g^{\mu\nu}\partial_\nu$ and $g=-\det(g_{\mu\nu})$. For the cosmon potential we assume $V=\mu^2\chi^2$ for large $\chi$ and $V=\lambda\chi^4$ for small $\chi$, as implemented by 
\be\label{2}
V=\frac{\mu^2\chi^4}{m^2+\chi^2}~,~\lambda=\frac{\mu^2}{m^2}. 
\ee
Stability requires $\alpha^2>0$, with $\alpha\to\infty$ corresponding to the ``conformal value''. The action \eqref{1} involves two dimensionless parameters $\alpha$ and $\lambda$. It specifies our model combined with an assumption on the $\chi$-dependence of particle masses that we discuss next. For this simple model we will find solutions of the homogeneous and isotropic field equations which have no singularity and can account for all present observations in cosmology.

Our model is based on the assumption of the existence of two fixed points for quantum gravity. For the first one, relevant for $\chi=0$, scale symmetry is exact and not spontaneously broken. All particles are massless. The second fixed point corresponds to $\chi\to\infty$ where scale symmetry is again exact. For $\chi\neq 0$ scale symmetry is spontaneously broken, however, resulting in massive particles. For $\chi\to\infty$ spontaneous scale symmetry breaking induces a Goldstone boson, the dilaton. Cosmology describes the transition between the two fixed points, with $\chi\to 0$ in the infinite past and $\chi\to\infty$ in the infinite future. Intermediate values of $\chi$ are associated to a crossover between the two fixed points. In this region scale symmetry is violated by the appearance of parameters with dimension mass or length. In the scalar-gravity sector this concerns the potential \eqref{2}.

Scale symmetry (or dilatation symmetry) plays a central role for the deeper particle physics understanding of our model and the judgment of its naturalness. (This symmetry is no longer easily visible in the Einstein frame.) Dilatation symmetry states the invariance of physics under a multiplicative scaling of all mass and associated length scales. It is realized as an exact symmetry if the quantum effective action contains no parameter with dimension of mass or length. The parameters $\mu$ or $m$ in the potential \eqref{2} have dimension mass and reflect a violation of scale symmetry (dilatation anomaly). Nevertheless, scale symmetry of the effective action \eqref{1} and the associated field equations is realized for the limits $\chi\to 0$ and $\chi\to\infty$. Besides the ``explicit scale symmetry breaking'' by the mass scales $\mu$ and $m$ any cosmological solution with a nonvanishing $\chi$ amounts to ``spontaneous scale symmetry breaking''. For present cosmology this spontaneous symmetry breaking is the dominant ingredient for the observed 
particle masses and the 
gravitational constant \cite{CW3}.

In general, quantum effects violate scale symmetry. This is reflected in the $\chi$-dependence of dimensionless couplings as gauge couplings $g$ or Yukawa couplings. For ``running couplings'' or non-vanishing $\beta$-functions, as $\tilde \beta_g=\chi\partial g(\chi)/\partial\chi$, the solution $g(\chi)$ can only depend on a dimensionless quantity as $\chi/m$ and therefore necessarily involves a mass scale $m$ (dimensional transmutation). By the same argument, any $\chi$-dependence of dimensionless ratios, as $m_e(\chi)/m_p(\chi),m_p(\chi)/\chi$ or $V(\chi)/\chi^4$, reflects a violation of scale symmetry. 

Since  a dimensionless quantity as $v=V/\chi^4$ can only depend on $m/\chi$ its flow equations in dependence on the ``renormalization scale'' $m$ is directly related to the flow equation in dependence on the field $\chi$
\be\label{3A}
m\frac{\partial v}{\partial m}=-\chi\frac{\partial v}{\partial\chi}=-\tilde \beta_v.
\ee
We will assume the existence of two fixed points for $m=0$ and $m\to\infty$, or correspondingly for $\chi\to\infty$ and $\chi=0$, $\tilde \beta_v(\chi=0)=0,~\tilde\beta_v(\chi\to\infty)=0$, with fixed point values $v(\chi=0)=\lambda,~v(\chi\to\infty)=0$. For a fixed point in the flow of all dimensionless couplings and ratios scale symmetry becomes exact. (This is well known from critical phenomena in statistical physics.) At a fixed point all $\beta$-functions for appropriately renormalized dimensionless quantities vanish. Since the $\beta$-functions and therefore their zeros are connected to quantum effects, we may call the scale symmetry associated to a fixed point ``quantum scale symmetry''. 

For $\chi\to 0$ we approximate in eq. \eqref{2} $m^2+\chi^2$ by $m^2$. The potential involves then only the dimensionless parameter $\lambda$ and becomes indeed scale invariant. (Scale symmetry breaking terms are suppressed by $\chi^2/m^2$.) In the asymptotic past $t\to-\infty$ the field $\chi$ approaches zero and our model realizes dilatation symmetry. On the other hand, for $\chi\to \infty$ the potential divided by the fourth power of the effective Planck mass goes to zero, $V/\chi^4\to \mu^2/\chi^2\to 0$. Up to small corrections $\sim \mu^2/\chi^2$ the limit $\chi\to\infty$ describes again the approach to a situation with effective quantum scale symmetry. A fixed point $\lim_{\chi\to\infty}(V/\chi^4)=0$ solves the cosmological constant problem if the cosmological solution implies that $\chi$ diverges for asymptotic time $t\to\infty$. This is realized for our model and explains why no fine tuning of parameters is needed in order to realize the tiny present dark energy density in units of the Planck mass. 

Finite non-vanishing values of $\chi$ describe a ``crossover region'' between the two fixed points. The crossover behavior is characterized by the presence of explicit mass scales, for our model given by $\mu$ and a second scale $m\approx 10^6\mu$. These mass scales describe the violation of dilatation symmetry in the scalar-gravity sector. The scalar-gravity sector of our model belongs to the class of models proposed by functional renormalization group investigations in dilaton quantum gravity \cite{HPRW}. There scale symmetry is a crucial ingredient for the establishment of a non-perturbatively renormalizable quantum field theory of gravity (asymptotic safety) \cite{Wei,Rev,Per}. 

Our model assumes that a crossover between two fixed points at $\chi=0$ and $\chi\to\infty$ is also realized in the particle physics sector. In these two limits the $\beta$-functions for all dimensionless quantities vanish. Dimensionless quantities become then independent of $\chi$ and all masses scale proportional to $\chi$. The values of the dimensionless quantities are different for the two fixed points, however. We emphasize that the running of couplings as the gauge coupling $g$ with $\chi/m$ should not be confounded with the running according to the standard model $\beta$-functions. The latter describe the dependence of $g$ on scale ratios, as momentum over Planck scale, or in our setting momentum$/\chi$. (In a Coleman-Weinberg setting they account for the dependence on the ratio Fermi scale/Planck scale.) In contrast, $\tilde \beta_g$ accounts for the simultaneous change of all scales (typically with fixed scale ratios close to fixed points) with respect to a reference scale $m$. A vanishing of $\tilde\beta_g$ simply means that $g$, normalized at momentum $\sim \chi$, does not depend on $\chi$. In consequence, $\Lambda_{QCD}$ will be proportional to $\chi$ such that $m_p\sim\chi$, as appropriate for a fixed point. (For more details see ref. \cite{CWEXP,CI}.) 

The departure from the fixed point at $\chi=0$ for increasing $\chi$ is typically characterized by a certain number of relevant or marginal parameters. They determine the scales where the crossover to the fixed point for $\chi\to\infty$ takes place. In case of marginal couplings these scales can be largely separated since the running is only logarithmic. We will assume that the crossover scale for the non-singlet sector of the standard model is common to all its dimensionless couplings since the $\beta$-functions connect the different couplings. For simplicity we identify the associated crossover scale with $\chi=m$. For the $SU(3)\times SU(2)\times U(1)$-singlet sector which describes physics beyond the standard model and influences the neutrino masses we assume a different crossover scale for $\chi$ near $M$ \cite{VG}. 

We will find that ``late cosmology'' after inflation corresponds to large values $\chi\gg m$, a region already close to the fixed point for $\chi\to\infty$. This can explain why the dimensionless gauge couplings of the standard model (normalized at a momentum scale $\chi$) are very close to their fixed point values, such that their variation with $\chi$ is tiny for late cosmology \cite{CWcross,CWcoupl}. In turn, the confinement scale of QCD scales proportional to $\chi$. The same argument can explain why the expectation value of the Higgs doublet scales $\kl \tilde h\kr\sim\chi$, and why Yukawa couplings become independent of $\chi$. Therefore the electron and quark masses, as well as the nucleon masses or nuclear binding energies, scale $\sim \chi$ with high accuracy. These features are crucial for a realistic model. Otherwise $m_e/M$ or $m_p/M$ would depend on the value of a time-variable scalar field in the Einstein frame, violating strict observational bounds from the time variation of fundamental 
couplings and tests of the equivalence principle. 

For the large singlet scale $M_{B-L}$, which characterizes the breaking of $B-L$ symmetry in the standard model singlet sector, we assume that the crossover takes place in the present cosmological epoch. This crossover typically describes a change from a fixed point value $M_{B-L}/\chi$ close to one for $\chi=0$ to a different fixed point value $M_{B-L}/\chi\ll 1$ for $\chi\to\infty$. The masses of neutrinos involve, besides the square of the vacuum expectation value of the Higgs-doublet $\tilde h$, the inverse of a high mass scale $M_{B-L}$ (see-saw mechanism), $m_\nu\sim\kl \tilde h\kr^2/M_{B-L}$. While $\kl \tilde h\kr\sim\chi$, our assumption implies that $M_{B-L}$ decreases for increasing $\chi$ in the present cosmological epoch.  This will trigger a transition from matter domination to dark energy domination once 
neutrinos become non-relativistic \cite{ABW,CWNEU} (typically 
at redshift $z\approx 5$). More quantitatively, we take the neutrino mass $m_\nu(\chi)$ as the average over neutrino species and define
\be\label{2A}
\tilde\gamma(\chi)=\frac12\frac{\partial\ln (m_\nu/\chi)}{\partial\ln\chi},
\ee
where $\tilde \gamma>0$ indicates the crossover behavior with $m_\nu$ increasing faster than $\chi$. In the future, for $\chi$ larger than the present value, neutrino masses may reach a fixed point scaling, similar to the other particle masses, $\tilde\gamma(\chi\to\infty)=0$. This is not relevant for the present and past, but decisive for the future of our universe. (For more details on the $\chi$-dependence of neutrino masses see ref. \cite{VG}.)

We should stress that the relevant flow equations for the dependence of couplings on $\chi/m$ have so far not be computed. Their computation should proceed along the lines outlined in ref. \cite{HPRW}, but this technically demanding task is outside the scope of this note. For the time being, we may simply summarize the main assumption of the crossover model for particle physics and their role for the cosmological evolution. The flow equation for the dependence of dimensionless couplings on $\chi/m$ describe a crossover between two fixed points as a scalar field $\chi$ varies between the extreme values $\chi\to 0$ and $\chi\to\infty$. The presence of several ``relevant'' or ``marginal'' couplings for the vicinity of the fixed point at $\chi=0$ can be associated to different characteristic mass scales for the crossover. Our model contains two such ``crossover scales''. As $\chi$ increases beyond $m$ the renormalizable couplings of the standard model switch to their asymptotic values for $\chi\to\infty$. Non-
renormalizable couplings reflecting a superheavy singlet sector, as neutrino mass terms, are supposed to change towards asymptotic values only in a region of larger $\chi$, with ongoing crossover behavior at the present value of $\chi$. In consequence, the resulting ``crossover cosmology'' also proceeds in two stages. A first stage for $\chi\approx m$ can be associated roughly with the end of inflation. The second stage of the crossover induces a substantial $\tilde\gamma$ for the present range of $\chi$. In turn, neutrinos stop the evolution of the cosmon as soon as they become non-relativistic, triggering the transition to dark energy domination. 

To sum up, the cosmology of our model depends on only three dimensionless parameters, $\mu/m,\alpha$ and the present value $\tilde \gamma=\tilde\gamma(\chi_0)$. (The parameter $\mu$, cf. eq. \eqref{1}, only sets the units of mass.) This simple setting can account for all presently available cosmological observations, covering the generation of primordial density fluctuations during inflation, the radiation- and matter-dominated periods with a small amount of early dark energy, and finally a present dark energy with a phenomenology close to a cosmological constant. Extensions of this minimal model with additional parameters will be discussed at later stages. 

{\em Field equations.} We assume a Robertson-Walker metric with scale factor $a(t)$ and vanishing spatial curvature, as well as a homogeneous $\chi(t)$. The time evolution of $\chi$ is governed by the scalar field equation which follows from variation of the effective action \eqref{1}. Besides the gradient of the potential $\partial V/\partial\chi$ it is driven by the coupling to the curvature scalar $R$. As a consequence, $\chi$ is found to increase with time, despite the minimum of the potential at $\chi=0$. (The solution with $\chi=0$ is unstable with respect to small deviations.) Inserting $R$ according to the gravitational field equations and defining 
\be\label{4}
s=\ln\left(\frac\chi m\right)~,~x=\frac{\chi^2}{m^2}=e^{2s},
\ee
the time evolution of the scalar field obeys
\be\label{5}
\ddot{s}+3H\dot{s}+2\dot{s}^2=
\frac{\alpha^2\mu^2x^2}{2(1+x)^2}
+\frac{\alpha^2g}{4}.
\ee
The Hubble parameter (for a standard Robertson-Walker metric) is determined by the gravitational field equation
\be\label{6}
(H+\dot{s})^2=
\frac{\mu^2x}{3(1+x)}
+\frac{2\dot{s}^2}{3\alpha^2}
+\frac{T_{00}}{3m^2x}.
\ee
Here $T_{\mu\nu}$ is the energy momentum tensor of radiation and matter and
\be\label{7}
g=\frac{q_\chi}{\chi}-\frac{T^\mu_\mu}{\chi^2}
\ee
involves $q_\chi$, the incoherent contribution of matter to the scalar field equation . 

For a species of particles with mass scaling proportional to $\chi^{2\tilde\gamma+1}$ one has $\chi q_\chi=-(2\tilde\gamma+1)(\rho-3p)$, with $\rho$ and $p$ the energy density and pressure of this species. (For a more detailed display and discussion of the field equations and more details on $q_\chi$ cf. ref. \cite{VG}.)  For massless particles both $q_\chi$ and $T^\mu_\mu$ vanish such that $g=0$ for the radiation dominated epoch. For particle masses $\sim\chi$ the two terms in eq. \eqref{7} cancel. As a consequence, $g$ obtains only a contribution from neutrinos $\sim\tilde\gamma$. It can be neglected as long as neutrinos are relativistic, i.e. for most of the cosmological evolution except a rather recent epoch.

{\em Slow freeze primordial cosmology.} For primordial cosmology we can neglect $T_{\mu\nu}$ and $g$. The scale $m$ no longer appears in the eqs. \eqref{5}, \eqref{6} for the evolution of $s$ and $H$, such that solutions for primordial cosmology will not involve the parameter $\lambda$ explicitly.  The ``beginning of the universe'' (the usual ``big bang'') corresponds in our setting to $\chi\to 0 ~(x\to 0)$ and we take the approximation $x\ll 1$. . Eqs. \eqref{5}, \eqref{6} admit a solution that behaves for $t\to-\infty$ in leading order as
\ba\label{8}
x=\left(\frac{-2}{\sqrt{3}\alpha^2\mu t}\right)^{\frac23}~,~H=\mu\sqrt{\frac x3}
~,~\dot{s}=-\frac{1}{3t}.
\ea
(More precisely, we can extend the solution \eqref{8} to a family of solutions by replacing $t\to t+c$ with constant $c$.) The asymptotic solution \eqref{8} is regular for arbitrarily large negative $t$ - our model has no big bang singularity. 

For $t\to-\infty$ the cosmon field $\chi$ goes to zero. The same holds for the Hubble parameter. The curvature scalar vanishes in the infinite past,
\be\label{9}
\lim_{t\to-\infty}R=4\mu^2x=4\lambda \chi^2\sim (-t)^{-\frac23},
\ee
where we use 
\be\label{10}
\lim_{t\to-\infty}
\left(\frac{\dot{H}}{H^2}\right)=
\lim_{t\to-\infty}
\left(\frac{\dot{s}}{H}\right)=
\left(\frac{-\alpha}{\sqrt{6}\mu t}\right)^{\frac23}
\to 0. 
\ee
By virtue of eq. \eqref{10} the leading order behavior of the curvature tensor for $t\to-\infty$ becomes
\be\label{11}
R_{\mu\nu\rho\sigma}=H^2
(g_{\mu\rho}g_{\nu\sigma}-
g_{\mu\sigma}g_{\nu\rho}).
\ee
All invariants formed by contracting powers of $R_{\mu\nu\sigma\lambda}$ and its covariant derivatives with powers of $g^{\mu\nu}$ vanish for $t\to-\infty$. The universe has evolved from an infinite past for which the geometry of space-time was locally flat and $\chi$ was infinitesimally close to zero.

The characteristic time for the expansion is given by the inverse Hubble parameter, diverging for $t\to-\infty$ as $H^{-1}\sim (-t)^{1/3}$. This is consistent with the notion of an extremely slow beginning. We will see that the order of magnitude of the Hubble parameter is always given by the first term in eq. \eqref{6} , $H^2\sim \mu^2 x/(1+x)$. It never exceeds substantially the present value $H\sim \mu$, which obtains for $x\gg 1$. For primordial cosmology $H$ is even further suppressed by a factor $\sqrt{x}$. (For another proposal of a slow evolution during inflation see ref. \cite{YP}.)

\medskip\noindent
{\em Eternal universe.} The ``physical time'' that has elapsed since the ``infinite past'' $t\to-\infty$ is infinite. In this sense the universe is eternal. For the solution \eqref{8} the scale factor increases 
\be\label{12A}
a(t)=a_0\exp \Bigg\{ -\left(-\frac{\sqrt{3}\mu t}{2\alpha}\right)^{\frac23}\Bigg\},
\ee
less fast than exponential, but faster than any power. Primordial cosmology describes an inflationary epoch. For $-t\gg 2\alpha/(\sqrt{3}\mu)$ conformal time $\big (d\eta=dt/a(t)\big)$ can be approximated by
\be\label{12B}
\eta(t)=-\left(-\frac{9\alpha^2 t}{2\mu^2}\right)^{\frac13}a^{-1}(t)
=-\frac{\alpha\sqrt{3}}{2\mu a}\ln\frac{a_0}{a}
\ee
(with possible additive constants $t\to t+c,\eta\to\eta+c')$. Between $t_0$ and $t_1$ photons travel a comoving distance $\eta(t_1)-\eta(t_0)$. For any given fixed time $t_1$ and $t_0\to-\infty$ this distance tends to infinity. For $\eta\to-\infty$ the scale factor obeys approximately
\be\label{14AB}
a(\eta)=-\frac{\alpha\sqrt{3}}{2\mu\eta}\ln
\left(-\frac{2\mu a_0\eta}{\alpha\sqrt{3}}\right).
\ee
In the past the causal structure of the universe is the same as for flat Minkowski space, with no singularity encountered. Geodesics are complete for photons or massive particles at rest (in comoving coordinates). In general, massive particles do not move on geodesics, due to an additional ``force'' induced by the change of mass. 

For massive particles with non-zero momentum $p(t)$ at finite time $t$ the momentum diverges for $t\to-\infty$. Such particles become photon-like in the infinite past. Proper time in units of the inverse particle mass, $d\tilde\tau=\chi d\tau$, can no longer  be used for a definition of physical time. Dimensionless time intervals $d\tilde\tau$ go to zero simply because the unit $\chi^{-1}$ diverges. One finds a finite distance in $\tilde\tau$ to the infinite past which reflects this shortcoming. For photons or particles that become photon-like in the infinite past we may use the number of oscillations of the wave function as a coordinate- and frame-invariant definition of physical time. Towards the infinite past this physical time is proportional to conformal time $\eta$, such that the physical time distance to the infinite past indeed diverges. (See ref. \cite{CWE} for more details on this issue.)

The family of cosmological solutions with asymptotic behavior \eqref{8} $(t\to t+c)$ is a stable attractor as time increases. This stability property defines the arrow of time \cite{VG}. We may insert eq. \eqref{6} into eq. \eqref{5} and rescale time, $\tilde t=\mu t,~s'=\partial s/\partial\tilde t=\dot s/\mu,~s''=\ddot{s}/\mu^2$, such that the evolution of $s$ is given by a simple non-linear second order differential equation which only involves the parameter $\alpha$,
\be\label{10A}
s''-s'^2+\sqrt{3}s'
\sqrt{\frac{e^{2s}}{1+e^{2s}}+\frac{2s'^2}{\alpha^2}}
-\frac{\alpha^2}{2}\frac{e^{4s}}{(1+e^{2s})^2}=0.
\ee
We have solved this equation numerically and find that all initial conditions with large negative $s(\tilde t_0)$ and small enough $s'(\tilde t_0)$ approach the family of scaling solutions as $t$ increases.

\begin{figure}[htb]
\begin{center}
\includegraphics[width=0.4\textwidth]{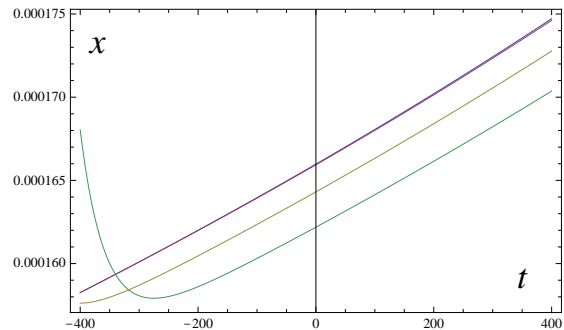}
\caption{Evolution of $x(t)$ for three different initial conditions at $t_0=-400/\mu$. The unit of time is $\mu^{-1}=10^{10}$yr and $\alpha=10$. We also display the asymptotic solution \eqref{8}. The upper curve shows the solution of eq. \eqref{10A} with the same initial conditions as the asymptotic solution. It cannot be distinguished from the asymptotic solution.}
\label{hbb_fig1}
\end{center}
\end{figure}

We display the evolution of $x$ for three different initial conditions in Fig. \ref{hbb_fig1}. All solutions reach soon the scaling solution \eqref{8} up to horizontal shifts $t\to t+c$. The upper curve coincides with the asymptotic solution \eqref{8} with $t\to t+\bar c$. Here we have chosen $\bar c$ such that $t=0$ corresponds to the time when primordial fluctuations on scales of cosmic structures have left the horizon (see below). Units of time are $\mu^{-1}=10^{10}$yr, such that the range shown in Fig. \ref{hbb_fig1} spans $8\cdot 10^{12}$yr. The solutions follow the scaling solution with high accuracy until inflation ends at $t=5470\mu^{-1}\approx 5.5\cdot 10^{13}$yr, when $x$ starts to increase rapidly and $H$ starts to decrease. The slow evolution in primordial cosmology is underlined by the observation that the time between horizon crossing and the end of inflation amounts to about $4000$ times the age of the universe in the big bang picture.

While the scaling solution can be continued to $t\to-\infty$, there is actually no need that the ``initial state'' of the universe is given precisely by the scaling solution. Due to the attractive character of the family of scaling solutions one may envisage a rather wide range of possible initial states. For example, the ``initial universe'' may be described by quantum fluctuations of $\chi$ around $\chi=0$ in flat space. For fluctuations that are homogeneous enough the scale factor will start an inflationary expansion according to the scaling solution \eqref{12A}. Different ``initial regions'' may have different expansion histories, as in chaotic inflation \cite{Li,Inf5}. The field equations have an exact homogeneous solution $\chi=0, R=0$ which corresponds to the fixed point. Small deviations from this solution are unstable. 

With the small values of $\dot{H}/H^2$ and $\dot{s}/H$ in eq. \eqref{10}, the primordial cosmology can be associated with an epoch of single field slow roll inflation. Inflation ends once $\dot{H}$ becomes of the same order as $H^2$. For $\alpha\gg 1$ this happens in the region of small $x\approx 1/\alpha\ll 1$ (see below). One expects a subsequent epoch of entropy production and heating of the universe, resulting in a radiation dominated epoch. Details of the entropy production depend on particle  masses and couplings in the region of small $x$ which we do not model here explicitly. A discussion of various mechanisms for entropy production after inflation in this type of varying gravity models (in the Einstein frame) can be found in refs. \cite{CI,HMS}.

{\em Slow freeze cosmological history.} The end of inflation starts the crossover from $x\ll1$ to $x\gg 1$. The subsequent radiation, matter and dark energy dominated epochs correspond to ``late cosmology'' with $x\gg 1$. For late cosmology the potential takes a simple quadratic form $V=\mu^2\chi^2$ such that only two cosmological parameters $\alpha$ and $\tilde\gamma$ remain relevant. 

The potential approximated by $V=\mu^2\chi^2$ has been discussed in detail in ref. \cite{VG} (model A), and also the kinetic term coincides with ref. \cite{VG} for $x\gg 1$. We only briefly describe here the main features of cosmology for $x\gg1$. The universe shrinks in the radiation dominated epoch with a constant negative Hubble parameter
\be\label{10B}
H=-\frac\alpha 2\mu,
\ee
while the value of the cosmon field $\chi$ increases exponentially according to 
\be\label{10C}
\dot{s}=\frac{\dot\chi}{\chi}=\alpha\mu~,~\chi\sim\exp (\alpha\mu t).
\ee
Due to the shrinking of the universe with scale factor $a\sim 1/\sqrt{\chi}$ the energy density in radiation increases $\sim \chi^2$,
\be\label{10D}
\rho_r=3\left(\frac{\alpha^2}{4}-1\right)\mu^2\chi^2,
\ee
similar to the potential and kinetic energy in the homogeneous scalar field which obey 
\be\label{10E}
\rho_h=V+\frac{2}{\alpha^2}\dot\chi^2
=3\mu^2\chi^2.
\ee
This results in a constant fraction of early dark energy \cite{EDE,DR}
\be\label{10F}
\frac{\rho_h}{\rho_r+\rho_h}=\Omega_e=\frac{4}{\alpha^2}.
\ee

While the temperature increases during radiation domination, $T\sim(\rho_r)^{\frac14}\sim \sqrt{\chi}$, the particle masses increase even faster $\sim\chi$. The equilibrium number density of a given species gets strongly Boltzmann-suppressed once a particle mass exceeds $T$. With Fermi scale $\kl\tilde h\kr\sim\chi$ and $\Lambda_{QCD}\sim\chi$, as well as constant dimensionless couplings, the decay rates scale $\sim\chi$, and all cross sections and interaction rates scale with the power of $\chi$ corresponding to their dimension. As a consequence, nucleosynthesis proceeds as in usual cosmology, now triggered by nuclear binding energies and the neutron-proton mass difference exceeding the temperature as $\chi$ increases. The evolution of all dimensionless quantities is the same as in standard cosmology, once we measure time in units of the (decreasing) inverse nucleon mass. The resulting element abundancies are essentially the same as in standard cosmology. The only difference arises from the presence of a 
fraction 
of early dark energy \eqref{10F}. This acts similarly to the presence of an additional radiation component, resulting in a lower bound on $\alpha$ from nucleosynthesis \cite{CW3,CW2,BS,BHM}. Later on, protons and electrons combine to hydrogen once the atomic binding energy (increasing $\sim\chi$) exceeds the temperature $T\sim\sqrt{\chi}$. Up to small effects of early dark energy the quantitative properties of the CMB-emission are the same as in standard cosmology. The effect of early dark energy on the detailed distribution of CMB-anisotropies gives so far the strongest bound on $\alpha,\alpha \gtrsim 10$, \cite{A2a,A2b,Re,Sievers:2013wk,A2d,PL}. 

The ratio of matter to radiation energy density increases as $\rho_m/\rho_r\sim\chi a$, with $a\sim\chi^{-\frac12}$ during radiation domination $(Ta=const.)$. This triggers the transition to a matter dominated scaling solution once $\rho_m$ exceeds $\rho_r$, given again by a shrinking de Sitter universe
\be\label{10G}
H=-\frac{\alpha\mu}{3\sqrt{2}}~,
~\dot{s}=\frac{\alpha\mu}{\sqrt{2}}~,
~\rho_m=\frac23(\alpha^2-3)\mu^2\chi^2,
\ee
with a constant fraction of early dark energy $\Omega_e=3/\alpha^2$. Observations of redshifts of distant galaxies are explained by the size of atoms shrinking faster than the distance between galaxies \cite{CWU,Na1,Na2,Na3}, resulting in an increase of the relevant ratio $\sim a\chi$. 

In a rather recent cosmological epoch $(z\approx 5)$ the neutrinos become non-relativistic. For $\tilde\gamma\gg 1$ the increase of their mass faster than $\chi$ stops effectively the time evolution of the cosmon field due to $g\neq 0$ in eq. \eqref{5}. The dark energy density $\rho_h$ remains frozen at the value it had at this moment, relating it to the average neutrino mass. More precisely, the cosmological solution oscillates around a very slowly evolving ``average solution'' for which the r.h.s. of eq. \eqref{5} vanishes to a good approximation, $V=\tilde \gamma\rho_\nu$. This yields for the homogeneous dark energy density $\rho_h$ the interesting quantitative relation \cite{ABW}
\be\label{10H}
\rho_h^{\frac14}=1.27\left(\frac{\tilde \gamma m_\nu}{{\rm eV}}\right)^{\frac14}10^{-3}{\rm eV}.
\ee
(Present neutrino masses on earth may deviate from the value of $m_\nu$ according to the cosmological average solution, due to oscillations and a reduction factor for neutrinos inside large neutrino lumps \cite{GNQ10,GNQ5A}. Cosmological bounds on $m_\nu$ are modified due to the mass variation.) For low redshift $z\lesssim 5$ cosmology is very similar to the $\Lambda$CDM-model with an effective equation of state for dark energy (more precisely the coupled cosmon-neutrino fluid) very close to $-1$,
\be\label{20A}
w=-1+\frac{\Omega_\nu}{\Omega_h}=-1+\frac{m_\nu(t_0)}{12{\rm eV}}.
\ee
An important distinction to the $\Lambda$CDM-model is the clumping of the neutrino background on very larges scales which may render it observable \cite{GNQ2,GNQ8,GNQ9,GNQ10}. The parameter $\mu$ in eq. \eqref{AA1} obtains from the observed value of the present dark energy density $\sqrt{\rho_h}=(2\cdot 10^{-3}$eV$)^2\approx\sqrt{V}=\mu M$. This also fixes $\tilde\gamma m_\nu=6.15$eV.

{\em Einstein frame.} Cosmological observables involve only dimensionless quantities. They are independent of the choice of fields used for their description (field relativity). For a quantitative discussion of observables it is advantageous to choose a ``frame'' (choice of ``field-coordinates'') in which the Planck mass is constant. Except for neutrinos this also results in constant particle masses for late cosmology.

Performing a Weyl scaling to the Einstein frame the effective action \eqref{1} becomes 
\be\label{W1}
\Gamma=\int d^4x\sqrt{g'}
\left(-\frac{M^2}{2}R'+{\cal L}_{{\rm kin}}+V'\right\},
\ee
with 
\ba\label{W2}
V'=\frac{\mu^2M^4}{m^2+\chi^2}~,~
{\cal L}_{{\rm kin}}=
\frac{2M^2}{\alpha^2\chi^2}
\partial^\mu\chi\partial_\mu\chi.
\ea
(Primes denote the metric and other quantities in the Einstein frame.) The standard normalization of the scalar kinetic term in the Einstein frame is realized by
\ba\label{S1}
\sigma=\frac{2M}{\alpha}\ln 
\left(\frac\chi m\right)=\frac{2Ms}{\alpha}~,~
{\cal L}_{{\rm kin}}=\frac12\partial^\mu\sigma\partial_\mu\sigma.
\ea
The potential, 
\be\label{S2}
V'(\sigma)=\lambda M^4
\left[1+\exp \left(\frac{\alpha\sigma}{M}\right)\right]^{-1}=
\frac{\lambda M^4}{(1+x)},
\ee
is constant for large negative $\sigma~(x\ll1)$ and decreases exponentially with $\sigma$ for large positive $\sigma~(x\gg 1)$. Neutrino masses depend on $\sigma$ according to $\partial \ln m_\nu/\partial\sigma=-\beta/M$, $\beta=-\alpha\tilde\gamma$, while all other particle masses in the standard model take their known fixed values. 

{\em Inflation and primordial density fluctuations.} We next describe the inflationary cosmology of our model in the Einstein frame where the spectrum  of primordial density fluctuations can be computed in a standard way. For $x\ll 1$ we can approximate $H$ by a constant,
\be\label{S3}
H=\sqrt{\frac\lambda3}M~,~\ddot{\sigma}+\sqrt{3\lambda}M\dot{\sigma}+\frac{\partial V'}{\partial\sigma}=0,
\ee
with 
\be\label{S4}
\frac{\partial V}{\partial\sigma}=-\frac{\alpha\lambda M^3 x}{(1+x)^2}
\approx -\alpha \lambda M^3\exp \left(\frac{\alpha\sigma}{M}\right).
\ee
(Here $H$ and $t$ refer to the Einstein frame and differ from corresponding quantities in the slow freeze scheme.) For the primordial inflationary phase one can neglect $\ddot{\sigma}$ as compared to $3H\dot{\sigma}$. The solutions of eq. \eqref{S3} then read
\be\label{S5}
\sigma(t)=-\frac{M}{\alpha}\ln 
\left (c_\sigma-\alpha^2\sqrt{\frac\lambda3}Mt\right),
\ee
with $c_\sigma$ an integration constant. They approximate a family of attractor solutions in the space of general isotropic and homogeneous solutions of the field equations. The logarithmic increase of $-\sigma$ for $t\to-\infty$ reflects the approach to $x=0$ with an inverse power law
\be\label{S6}
x=\left[{c_\sigma-\alpha^2\sqrt{\frac\lambda3}Mt}\right]^{-1}.
\ee
This corresponds to the asymptotic solution \eqref{8}, taking into account that the time variable differs between two frames. 

We observe that also in the Einstein frame the solution \eqref{S6} can be continued to $t\to-\infty$, with finite curvature invariants in this limit. The absence of a big bang singularity is similar to the asymptotic de-Sitter space found earlier in higher dimensional inflation \cite{Inf5,HI3}. While de Sitter space shows no singularity in geometric invariants, it is sometimes considered as singular due to geodesic incompleteness \cite{BV,MV}. The proper time distance to the infinite past $t\to-\infty$ diverges for all ``asymptotic massive particles'' for which the ratio momentum/mass remains finite. In contrast, one finds a finite proper time distance (geodesic incompleteness) for particles that become photon-like in the infinite past. This is the behavior we have found already in the freeze frame, with dimensionless proper time $\tilde\tau$ in the freeze frame corresponding to proper time in the Einstein frame. Proper time cannot be used for a definition of physical time for photon-like particles. Thus the ``incompleteness'' of time-like geodesics does not indicate a ``beginning'' of the use of proper 
time. This issue is discussed in detail in ref. \cite{CWE}.

The inflationary phase ends for $-\dot{H}/H^2=\dot{\sigma}^2/(2M^2H^2)$ 

\noindent
$\approx 1$. With
\be\label{S7}
\frac{\dot{\sigma}}{MH}=\alpha x
\ee
we may associate the end of inflation with 
\be\label{S8}
x_f=\frac1\alpha.
\ee
(With $\ddot{\sigma}/(3H\dot{\sigma})\approx \alpha^2 x/3$ the validity of the neglection of the $\ddot{\sigma}$-term ends for  a somewhat smaller value $\bar x_f=3/\alpha^2$. Substituting $\bar x_f$ for $x_f$ only results in very small modifications of the estimates below. The time needed for the evolution from $\bar x_f$ to $x_f$ is less than an $e$-folding.)

A quantitative discussion of the generation of density fluctuations during the inflationary epoch is facilitated by the choice of a scalar field variable for which the potential takes a ``standard'' exponential form. Indeed, for 
\be\label{W3}
\varphi =\frac M\alpha\ln
\left(1+\frac{\chi^2}{m^2}\right),
\ee
the potential reads
\be\label{B2A}
V'=\lambda M^4\exp 
\left(-\frac{\alpha\varphi}{M}\right),
\ee
while the kinetic term involves now a non-trivial ``kinetial'' $k^2(\varphi)$,
\ba\label{W4}
{\cal L}_{{\rm kin}}&=&\frac12 k^2(\varphi)\partial^\mu\varphi\partial_\mu\varphi,\nn\\
k^2&=&
\left(\frac{m^2+\chi^2}{\chi^2}\right)^2=
\frac{\exp\left(\frac{2\alpha\varphi}{M}\right)}{\left(\exp\left(\frac{\alpha\varphi}{M}\right)-1\right)^2}.
\ea
For $\varphi\to\infty$ the field $\varphi$ has the standard normalization, $k^2(\varphi\to\infty)=1$ and our model describes a standard quintessence model with an exponentially decreasing potential. (In this limit the difference between $\varphi$ and $\sigma$ becomes insignificant.)

The inflationary period corresponds to $\chi\to 0$ or $\varphi\to 0$. With our normalization of $\varphi$ the computation of the properties of the primordial density fluctuations becomes very simple \cite{VG}. The slow roll parameters $\epsilon$ and $\eta$ depend only on the kinetial $k^2(\varphi)$ 
\be\label{I1}
\epsilon=\frac{\alpha^2}{2k^2}~,~\eta=2\epsilon-\frac M\alpha\frac{\partial\epsilon}{\partial\varphi}.
\ee
For $\varphi\to 0$ one has 
\be\label{I2}
k^{-2}=\frac{\alpha^2\varphi^2}{M^2}~,~\epsilon=\frac{\alpha^4\varphi^2}{2M^2}~,~\eta=-\frac{\alpha^3\varphi}{M}.
\ee
The slow roll conditions $\epsilon\ll 1~,~|\eta|\ll 1$ are obeyed for $\varphi\to 0$. Following the behavior at larger values of $\varphi$ we use 
\ba\label{I3}
\epsilon&=&\frac{\alpha^2x^2}{2(1+x)^2}~,~x=\exp \left(\frac{\alpha\varphi}{M}\right)-1,\nn\\
\eta&=&2\epsilon-(1+x)\frac{\partial\epsilon}{\partial x}=-\frac{\alpha^2x(1-x)}{(1+x)^2}.
\ea

The number of $e$-foldings before the end of inflation is related to $x$ by
\be\label{I4}
N(x)=\frac{1}{\alpha M}\int^{\varphi_f}_\varphi d\varphi'k^2(\varphi')
=\frac{1}{\alpha^2}\int^{x_f}_xdx'\frac{k^2(x')}{1+x'},
\ee
where $\varphi_f$ or $x_f$ denote the values at the end of inflation. Using eq. \eqref{W4}, i.e. $k^2=(1+x)^2/x^2$, one obtains
\ba\label{I5}
N(x)=\frac{c(x)}{\alpha^2x}~,~
c(x)=1-\frac{x}{x_f}+x\ln \frac{x_f}{x}.
\ea
The factor $c(x)$ is close to one for small $x/x_f$. For $\alpha^2\gg 1$ and $N\approx 60$ one finds indeed a very small
\be\label{I6}
x(N)=\frac{c}{N\alpha^2},
\ee
and eq. \eqref{S8} yields in leading order
\be\label{Ib}
c=1-\frac{1}{N\alpha}.
\ee
Up to tiny corrections this implies for the slow roll parameters $\eta$ and $\epsilon$
\be\label{I7}
\epsilon\approx \frac{1}{2 N^2\alpha^2}~,~\eta\approx-\frac{1}{N}~,~\epsilon\approx \frac{\eta^2}{2\alpha^2}.
\ee

Our crossover model predicts the spectral index to be independent of $\alpha$
\be\label{I8}
n=1+2\eta-6\epsilon\approx 1-\frac{2}{N}\approx 0.96-0.967,
\ee
where we have taken $50< N<60$. The amplitude ratio $r$ of tensor fluctuations compared to scalar fluctuations is found to be very small, depending on $\alpha$
\be\label{I9}
r=16\epsilon=\frac{8}{N^2\alpha^2} <3\cdot 10^{-5}.
\ee
Both values fit very well the findings of the Planck satellite \cite{PL}.
For reproducing the measured amplitude of the primordial density fluctuations we need
\be\label{I10}
24\pi^2\Delta=\frac{V'}{\epsilon M^4}\approx \frac{\mu^2}{\epsilon m^2}\approx 5\cdot 10^{-7}.
\ee
One infers that the ratio $\mu/m$ must be small, typically of the order $10^{-6}$ for $\alpha\approx 10$,
\be\label{I11}
\frac\mu m\approx \frac{5}{N\alpha}\cdot 10^{-4}.
\ee

The entropy production at the end of inflation depends on (possibly $\varphi$-dependent) masses and couplings of particles with mass smaller than $\sqrt{\lambda}M$, that we do not specify here, cf. refs. \cite{CI,HMS} The asymptotic regime of late cosmology $x\gg 1$ is reached at early stages of the radiation dominated epoch. We will assume that masses and couplings of all particles (except for the neutrino sector) are already close to their fixed point values for $\chi\to\infty$ long before nucleosynthesis.  They are then constant in the Einstein frame. The effective action \eqref{W1}, \eqref{B2A}, \eqref{W4} describes a standard model of dynamical dark energy or quintessence with an exponential potential. The cosmic attractor or ``tracker'' solution \cite{CW3,CW2,QU4} results in a decrease of dark energy proportional to radiation or matter. This explains the present tiny value of the dark energy density as a consequence of the large age of the universe. Neutrinos couple to $\varphi$ 
with $\beta=-\tilde \gamma\alpha$, 
stopping the evolution of $\varphi$ once they get non-relativistic according to the ``growing neutrino quintessence'' scenario \cite{ABW,CWNEU}. 

{\em Future of the universe}. The future of the universe depends on the behavior of the neutrino mass as $\varphi$ or $\chi$ increase. We will assume here that the crossover towards the fixed point for $\chi\to\infty$ ends for $\chi\gg \bar m$. For $\chi\gg \bar m$ neutrino masses scale $\sim \chi$ in the freeze scheme and become independent of $\varphi$ in the Einstein frame, $\tilde \gamma\to 0, \beta\to 0$. As a consequence, cosmology turns back to a matter dominated universe. In the far future the role of dark energy will again be reduced to a small fraction $\Omega_h=3/\alpha^2$ according to the scaling solution. In the Einstein frame the neutrino mass will reach a fixed value $\bar m_\nu$. For sufficiently large $\bar m_\nu$, (say $\bar m_\nu>1$keV) the neutrinos will by far dominate the energy density of the universe. 

In the freeze scheme the masses of all particles, including now the neutrinos, continue to increase $\sim\chi$. The future universe will shrink again, according to the solution \eqref{10G}. The scale factor will go exponentially to zero
\be\label{48X}
a(t)=\bar a\exp (-\bar h t)~,~\bar h=\frac{\alpha\mu}{3\sqrt{2}}.
\ee
In conformal time this implies (for suitable constants $\bar a,\bar \eta$)
\be\label{48Y}
a(\eta)=\frac{\bar a}{\bar h(\eta-\bar\eta)}~,~\eta=\bar\eta+\frac{1}{\bar h}\exp (\bar h t).
\ee
Photons can travel infinitely for $\eta\to\infty$. Also the trajectories of massive particles can be continued to infinite time, Between the (infinite) past for $\chi\to 0$ and the (infinite) future for $\chi\to\infty$ the universe is regular. 

{\em Discussion.} We have described the cosmology of our crossover model in two different pictures or frames, as a slow freeze or a big bang, according to the choice of frame \eqref{1} or \eqref{W1}. One may ask which picture is more ``natural''. In a situation where particle masses and the cosmic time scale $H^{-1}$ evolve differently with time, this amounts to a gauge of clocks in units of inverse particle masses (Einstein frame) or in cosmological time units (freeze scheme). For late cosmology the use of time units inversely proportional to the particle masses seems preferable and one may opt for the Einstein frame as the most natural picture. For primordial cosmology the particle masses approach zero for $t\to-\infty$ in the freeze scheme. This is no problem, massless particles being fully consistent. A choice of time scale given by the inverse of the particle masses may no longer be well adapted, however. It seems more natural to use a fixed ``cosmological time unit'', as given, for 
example, by the parameter 
$\mu^{-1}$. One may then prefer the freeze scheme \eqref{1}, with characteristic time scales of the cosmic evolution given for all epochs by $\mu^{-1}=10^{10}$yr, cf. eqs. \eqref{10B}, \eqref{10G}. This choice avoids the problem of characteristic times getting shorter and shorter as we go back in the history of the 
universe in the Einstein frame. In this sense many apparently ``problematic aspects'' of the big bang picture can be associated to a choice of clocks that is not optimally suited. We emphasize that a choice of clocks does not only involve a coordinate choice, but also a choice of the metric or frame. 

In the freeze picture Newton's ``constant'' diverges $\sim \chi^{-2}$ for $t\to-\infty$ and one may be worried about gravity getting strong and uncontrollable. For particle masses $\sim\chi$, as appropriate for a fixed point for $\chi\to 0$, the strength of the long-distance gravitational attraction between two particles becomes independent of $\chi$, however. The propagation of massless particles ``feels'' only the conformal aspects of geometry.  These two effects of gravity are actually independent of $\chi$ and do not diverge for $\chi\to 0$. For a discussion of graviton-graviton scattering we add to the effective action a term $\sim C\sqrt{g}R^2$. With dimensionless $C$ this term is scale invariant and its presence is expected at the fixed point $\chi=0$. We have neglected this term so far since it plays at all stages a subleading role for the cosmological solution if $\lambda=\mu^2/m^2\ll 1$. In the freeze frame one finds indeed $R^2\ll\chi^2|R|$ due to $H^2/\chi^2=\lambda/3$. Graviton scattering with fixed non-zero momentum will be dominated, however, by the $R^2$-type term in the limit $\chi\to0$. In this limit the gravitational interactions are governed by fourth order gravity and become independent of $\chi$, cf. ref. \cite{CWE}. Thus graviton-graviton scattering or the gravitational scattering of particles at non-zero momentum transfer do not become divergent in the infinite past. 

For a given effective action all frames can be used equivalently. The choice of frame becomes important, however, if one wants to relate an effective action of the type \eqref{1} or \eqref{W1} to a quantum computation of a possible (non-perturbative) ultraviolet fixed point which can define consistent quantum gravity \cite{HPRW}. In principle, the effect of quantum fluctuations can be computed equivalently in arbitrary frames. The transition between two frames involves, however, a Jacobian from the functional measure, which is often difficult to handle. In practice, one will select a given frame by assuming (implicitly) a ``unit Jacobian''. 

In this context we note that a judgment of the naturalness of a given effective action in a quantum field theory of gravity gets possibly obscured by the fact that the effective action looks rather different in different frames. For example, an appropriate $\chi$-dependent rescaling of the metric can cast the effective action \eqref{1} into the form 
\ba\label{36}
\Gamma&=&\int d^4 x\sqrt{\tilde g}
\left\{-
\frac{\chi^2+m^2}{2}\tilde R\right.\nn\\
&&\left.+\frac12\tilde K\partial^\mu\chi\partial_\mu\chi+\mu^2
(m^2+\chi^2)\right\},
\ea
with 
\be\label{37}
\tilde K=\frac{4(\chi^2+m^2)}{\alpha^2\chi^2}-\frac{6\chi^2}{\chi^2+m^2}.
\ee
In this frame the effective Planck mass and the cosmon potential reach for $\chi\to 0$ constant values $m$ and $V_0=\mu^2 m^2=\lambda m^4$. The potential $\tilde V$ is now a simple polynomial, while $\tilde K$ diverges $\sim \chi^{-2}$. The scale symmetry for $\chi\to 0$ is no longer visible. Late cosmology is described in the same freeze picture as for the action \eqref{1}, only inflation is described in a different picture. 

Cosmological predictions are rather insensitive to many details of models similar to eq. \eqref{1}. For example, the existence of scaling solutions with a constant early dark energy fraction $\Omega_e$ (and corresponding decrease of the dark energy density $\sim t^{-2}$ in the Einstein frame) only requires that $V$ increases for large $\chi$ with a power less than four \cite{CW3}. Replacing in eq. \eqref{1} the potential by 
\be\label{38}
V=\frac{\lambda m^4\chi^4}{(m^2+\chi^2)^2}
\ee
and rescaling $\alpha\to\alpha/2$ leaves all predictions for late cosmology $(x\gg 1)$ unchanged \cite{VG}. Now $V$ becomes constant for large $\chi,V_0=\lambda m^4=(2\cdot 10^{-3}$eV)$^4$. Also the spectrum of primordial density fluctuations generated during inflation remains almost the same, up to a change $\alpha\to\alpha/2$ in eqs. \eqref{I9}, \eqref{I11}. (More precisely, the expression $k^2=(1+x)^2/x^2$ remains the same, but now $1+x=\exp(\alpha\varphi/2M),~\eta\approx -\alpha^2x/2,~x(N)\approx 2/(N\alpha^2)$. While $n$ remains unchanged, one has $r=32/(N^2\alpha^2),~\sqrt{\lambda}=10^{-3}/(N\alpha)$.) Nevertheless, the picture of the universe is different from the slow freeze picture with potential \eqref{2}. One now finds a static universe during radiation domination and expansion during matter domination, similar to late cosmology for model (B) in ref. \cite{VG}. With $\lambda=10^{-12}$ the dimensionful parameters are given by $m=2$eV, $\mu=\sqrt{\lambda}m=2\cdot 10^{-6}$eV, differing strongly from 
eq. \eqref{AA1}.

As mentioned before, we could add to the effective action \eqref{1} terms $\sim R^2,R_{\mu\nu}R^{\mu\nu}$ or $R_{\mu\nu\rho\sigma}R^{\mu\nu\rho\sigma}$ with constant coefficients (or coefficients showing a crossover between different constants for the fixed points at $\chi\to 0$ and $\chi\to\infty$). Such terms are compatible with our crossover model with scale symmetry for $\chi\to 0$ and $\chi \to \infty$. Comparing $R$ with $\chi^2$ for the primordial scaling solution \eqref{8}, \eqref{9} yields $R/\chi^2=4\lambda$. In view of the small value for $\lambda=\mu^2/m^2$ required from the amplitude of primordial density fluctuations \eqref{I11}, modifications from higher order curvature terms are tiny unless their coefficients are huge. The role of such terms for late cosmology is even suppressed further.

Of course, one may also consider modifications of the potential or kinetic term in the effective action \eqref{1} that have more pronounced effects on the spectrum of the primordial density fluctuation. For example, replacing eq. \eqref{2} by \cite{CWE}
\be\label{ZA}
V=\frac{\mu^2\chi^2 x}{1+x}~,~x=\left(\frac{\chi^2}{m^2}\right)^{1-\frac{\tilde\alpha}{\alpha}}
\ee
yields for the spectral index $n$ and tensor amplitude $r$
\ba\label{43}
n&=&1-2\alpha^2x(N)-\tilde\alpha^2~,~r=8(\tilde\alpha+\alpha x(N))^2,\nn\\
x(N)&=&(\tilde\alpha/\alpha)\big(\exp\big\{\tilde\alpha(\alpha-\tilde\alpha)N\big\}-1\big)^{-1}. 
\ea
For $\tilde\alpha\alpha\gg1/N$ this implies $r=8\tilde\alpha^2$, $n=1-r/8$. If substantial tensor fluctuations are observed one may need the additional parameter $\tilde\alpha$, but for the moment we may stick to the simplest model withe $\tilde\alpha=0$.

We conclude that in the freeze picture rather simple and minimal models can describe a realistic cosmology from inflation to present dark energy domination. Our crossover model with the three parameters $\alpha,\lambda$ and $\tilde \gamma$ has no more free parameters than the standard $\Lambda$CDM model. The parameter $\tilde \gamma$ determines the present dark energy density, cf. eq.\eqref{10H}, and the combination $\alpha\sqrt{\lambda}$ is fixed by the amplitude of the primordial density fluctuations, cf. eq. \eqref{I11}. The spectral index $n$ is not a free parameter, in contrast to $\Lambda$CDM. This allows for falsification by a precision measurement of $n$. On the other hand, the parameter $\alpha$ has no correspondence in the $\Lambda$CDM-model. It determines the evolution of the scalar field for all cosmological epochs. Both the prediction of the tensor to scalar ratio $r$ in eq. \eqref{I9} and of the fraction of early dark energy in eq. \eqref{10F} (and similar for matter domination) 
depend on $\alpha$. This relates quantities in late and early cosmology, namely the fraction in early dark energy $\Omega_e$ and the tensor amplitude $r$. Our model predicts $n\approx 0.96$ and $r<3\cdot 10^{-5}$.  Finding nonvanishing early dark energy or the large neutrino lumps of growing neutrino quintessence could be interpreted as important hints towards our scenario. A cosmological model with no more free parameters than the $\Lambda$CDM-model is highly predictive!

\vspace{2.0cm}\noindent

\bibliography{hot_big_bang_or_slow_freeze}

\begin{thebibliography}{49}
\expandafter\ifx\csname natexlab\endcsname\relax\def\natexlab#1{#1}\fi
\expandafter\ifx\csname bibnamefont\endcsname\relax
  \def\bibnamefont#1{#1}\fi
\expandafter\ifx\csname bibfnamefont\endcsname\relax
  \def\bibfnamefont#1{#1}\fi
\expandafter\ifx\csname citenamefont\endcsname\relax
  \def\citenamefont#1{#1}\fi
\expandafter\ifx\csname url\endcsname\relax
  \def\url#1{\texttt{#1}}\fi
\expandafter\ifx\csname urlprefix\endcsname\relax\def\urlprefix{URL }\fi
\providecommand{\bibinfo}[2]{#2}
\providecommand{\eprint}[2][]{\url{#2}}

\bibitem[{\citenamefont{Wetterich}(2013{\natexlab{a}})}]{CWU}
\bibinfo{author}{\bibfnamefont{C.}~\bibnamefont{Wetterich}},
  \bibinfo{journal}{Dark Universe} \textbf{\bibinfo{volume}{2}},
  \bibinfo{pages}{184} (\bibinfo{year}{2013}{\natexlab{a}}),
  \eprint{1303.6878}.

\bibitem[{\citenamefont{Wetterich}(1988{\natexlab{a}})}]{CW3}
\bibinfo{author}{\bibfnamefont{C.}~\bibnamefont{Wetterich}},
  \bibinfo{journal}{Nucl.Phys.} \textbf{\bibinfo{volume}{B302}},
  \bibinfo{pages}{668} (\bibinfo{year}{1988}{\natexlab{a}}).

\bibitem[{\citenamefont{Ratra and Peebles}(1988)}]{RP}
\bibinfo{author}{\bibfnamefont{B.}~\bibnamefont{Ratra}} \bibnamefont{and}
  \bibinfo{author}{\bibfnamefont{P.}~\bibnamefont{Peebles}},
  \bibinfo{journal}{Phys.Rev.} \textbf{\bibinfo{volume}{D37}},
  \bibinfo{pages}{3406} (\bibinfo{year}{1988}).

\bibitem[{\citenamefont{Wetterich}(2013{\natexlab{b}})}]{CI}
\bibinfo{author}{\bibfnamefont{C.}~\bibnamefont{Wetterich}},
  \bibinfo{journal}{Phys.Lett.} \textbf{\bibinfo{volume}{B726}},
  \bibinfo{pages}{15} (\bibinfo{year}{2013}{\natexlab{b}}), \eprint{1303.4700}.

\bibitem[{\citenamefont{Weyl}(1918)}]{Weyl}
\bibinfo{author}{\bibfnamefont{H.}~\bibnamefont{Weyl}},
  \bibinfo{journal}{Sitzungsber. Preuss. Akad. Wiss. Berlin (Math.Phys.)}
  \textbf{\bibinfo{volume}{1918}}, \bibinfo{pages}{465} (\bibinfo{year}{1918}).

\bibitem[{\citenamefont{Dicke}(1962)}]{Di}
\bibinfo{author}{\bibfnamefont{R.}~\bibnamefont{Dicke}},
  \bibinfo{journal}{Phys.Rev.} \textbf{\bibinfo{volume}{125}},
  \bibinfo{pages}{2163} (\bibinfo{year}{1962}).

\bibitem[{\citenamefont{Wetterich}(1988{\natexlab{b}})}]{CW1}
\bibinfo{author}{\bibfnamefont{C.}~\bibnamefont{Wetterich}},
  \bibinfo{journal}{Nucl.Phys.} \textbf{\bibinfo{volume}{B302}},
  \bibinfo{pages}{645} (\bibinfo{year}{1988}{\natexlab{b}}).

\bibitem[{\citenamefont{Damour and Esposito-Farese}(1992)}]{DamE}
\bibinfo{author}{\bibfnamefont{T.}~\bibnamefont{Damour}} \bibnamefont{and}
  \bibinfo{author}{\bibfnamefont{G.}~\bibnamefont{Esposito-Farese}},
  \bibinfo{journal}{Class.Quant.Grav.} \textbf{\bibinfo{volume}{9}},
  \bibinfo{pages}{2093} (\bibinfo{year}{1992}).

\bibitem[{\citenamefont{Flanagan}(2004)}]{FR3}
\bibinfo{author}{\bibfnamefont{E.~E.} \bibnamefont{Flanagan}},
  \bibinfo{journal}{Class.Quant.Grav.} \textbf{\bibinfo{volume}{21}},
  \bibinfo{pages}{3817} (\bibinfo{year}{2004}), \eprint{gr-qc/0403063}.

\bibitem[{\citenamefont{Catena et~al.}(2007)\citenamefont{Catena, Pietroni, and
  Scarabello}}]{Cat1}
\bibinfo{author}{\bibfnamefont{R.}~\bibnamefont{Catena}},
  \bibinfo{author}{\bibfnamefont{M.}~\bibnamefont{Pietroni}}, \bibnamefont{and}
  \bibinfo{author}{\bibfnamefont{L.}~\bibnamefont{Scarabello}},
  \bibinfo{journal}{Phys.Rev.} \textbf{\bibinfo{volume}{D76}},
  \bibinfo{pages}{084039} (\bibinfo{year}{2007}), \eprint{astro-ph/0604492}.

\bibitem[{\citenamefont{Deruelle and Sasaki}(2011)}]{DeS}
\bibinfo{author}{\bibfnamefont{N.}~\bibnamefont{Deruelle}} \bibnamefont{and}
  \bibinfo{author}{\bibfnamefont{M.}~\bibnamefont{Sasaki}},
  \bibinfo{journal}{Springer Proc.Phys.} \textbf{\bibinfo{volume}{137}},
  \bibinfo{pages}{247} (\bibinfo{year}{2011}), \eprint{1007.3563}.

\bibitem[{\citenamefont{Henz et~al.}(2013)\citenamefont{Henz, Pawlowski,
  Rodigast, and Wetterich}}]{HPRW}
\bibinfo{author}{\bibfnamefont{T.}~\bibnamefont{Henz}},
  \bibinfo{author}{\bibfnamefont{J.~M.} \bibnamefont{Pawlowski}},
  \bibinfo{author}{\bibfnamefont{A.}~\bibnamefont{Rodigast}}, \bibnamefont{and}
  \bibinfo{author}{\bibfnamefont{C.}~\bibnamefont{Wetterich}},
  \bibinfo{journal}{Phys.Lett.} \textbf{\bibinfo{volume}{B727}},
  \bibinfo{pages}{298} (\bibinfo{year}{2013}), \eprint{1304.7743}.

\bibitem[{\citenamefont{Weinberg}(1979)}]{Wei}
\bibinfo{author}{\bibfnamefont{S.}~\bibnamefont{Weinberg}}
  (\bibinfo{year}{1979}), \eprint{General Relativity: An Einstein Centenary
  Survey, eds. S. W. Hawking and W. Israel, Cambridge University Press}.

\bibitem[{\citenamefont{Reuter}(1998)}]{Rev}
\bibinfo{author}{\bibfnamefont{M.}~\bibnamefont{Reuter}},
  \bibinfo{journal}{Phys.Rev.} \textbf{\bibinfo{volume}{D57}},
  \bibinfo{pages}{971} (\bibinfo{year}{1998}), \eprint{hep-th/9605030}.

\bibitem[{\citenamefont{Narain and Percacci}(2010)}]{Per}
\bibinfo{author}{\bibfnamefont{G.}~\bibnamefont{Narain}} \bibnamefont{and}
  \bibinfo{author}{\bibfnamefont{R.}~\bibnamefont{Percacci}},
  \bibinfo{journal}{Class.Quant.Grav.} \textbf{\bibinfo{volume}{27}},
  \bibinfo{pages}{075001} (\bibinfo{year}{2010}), \eprint{0911.0386}.

\bibitem[{\citenamefont{Wetterich}(2008)}]{CWEXP}
\bibinfo{author}{\bibfnamefont{C.}~\bibnamefont{Wetterich}},
  \bibinfo{journal}{Phys.Rev.} \textbf{\bibinfo{volume}{D77}},
  \bibinfo{pages}{103505} (\bibinfo{year}{2008}), \eprint{0801.3208}.

\bibitem[{\citenamefont{Wetterich}(2014{\natexlab{a}})}]{VG}
\bibinfo{author}{\bibfnamefont{C.}~\bibnamefont{Wetterich}},
  \bibinfo{journal}{Phys.Rev.} \textbf{\bibinfo{volume}{D89}},
  \bibinfo{pages}{024005} (\bibinfo{year}{2014}{\natexlab{a}}),
  \eprint{1308.1019}.

\bibitem[{\citenamefont{Wetterich}(2003{\natexlab{a}})}]{CWcross}
\bibinfo{author}{\bibfnamefont{C.}~\bibnamefont{Wetterich}},
  \bibinfo{journal}{Phys.Lett.} \textbf{\bibinfo{volume}{B561}},
  \bibinfo{pages}{10} (\bibinfo{year}{2003}{\natexlab{a}}),
  \eprint{hep-ph/0301261}.

\bibitem[{\citenamefont{Wetterich}(2003{\natexlab{b}})}]{CWcoupl}
\bibinfo{author}{\bibfnamefont{C.}~\bibnamefont{Wetterich}},
  \bibinfo{journal}{JCAP} \textbf{\bibinfo{volume}{0310}}, \bibinfo{pages}{002}
  (\bibinfo{year}{2003}{\natexlab{b}}), \eprint{hep-ph/0203266}.

\bibitem[{\citenamefont{Amendola et~al.}(2008)\citenamefont{Amendola, Baldi,
  and Wetterich}}]{ABW}
\bibinfo{author}{\bibfnamefont{L.}~\bibnamefont{Amendola}},
  \bibinfo{author}{\bibfnamefont{M.}~\bibnamefont{Baldi}}, \bibnamefont{and}
  \bibinfo{author}{\bibfnamefont{C.}~\bibnamefont{Wetterich}},
  \bibinfo{journal}{Phys.Rev.} \textbf{\bibinfo{volume}{D78}},
  \bibinfo{pages}{023015} (\bibinfo{year}{2008}), \eprint{0706.3064}.

\bibitem[{\citenamefont{Wetterich}(2007)}]{CWNEU}
\bibinfo{author}{\bibfnamefont{C.}~\bibnamefont{Wetterich}},
  \bibinfo{journal}{Phys.Lett.} \textbf{\bibinfo{volume}{B655}},
  \bibinfo{pages}{201} (\bibinfo{year}{2007}), \eprint{0706.4427}.

\bibitem[{\citenamefont{Piao}(2011)}]{YP}
\bibinfo{author}{\bibfnamefont{Y.-S.} \bibnamefont{Piao}}
  (\bibinfo{year}{2011}), \eprint{1112.3737}.

\bibitem[{\citenamefont{Wetterich}(2014{\natexlab{b}})}]{CWE}
\bibinfo{author}{\bibfnamefont{C.}~\bibnamefont{Wetterich}}
  (\bibinfo{year}{2014}{\natexlab{b}}), \eprint{1404.0535}.

\bibitem[{\citenamefont{Linde}(1983)}]{Li}
\bibinfo{author}{\bibfnamefont{A.~D.} \bibnamefont{Linde}},
  \bibinfo{journal}{Phys.Lett.} \textbf{\bibinfo{volume}{B129}},
  \bibinfo{pages}{177} (\bibinfo{year}{1983}).

\bibitem[{\citenamefont{Shafi and Wetterich}(1983)}]{Inf5}
\bibinfo{author}{\bibfnamefont{Q.}~\bibnamefont{Shafi}} \bibnamefont{and}
  \bibinfo{author}{\bibfnamefont{C.}~\bibnamefont{Wetterich}},
  \bibinfo{journal}{Phys.Lett.} \textbf{\bibinfo{volume}{B129}},
  \bibinfo{pages}{387} (\bibinfo{year}{1983}).

\bibitem[{\citenamefont{Hossain et~al.}(2014)\citenamefont{Hossain, Myrzakulov,
  Sami, and Saridakis}}]{HMS}
\bibinfo{author}{\bibfnamefont{M.~W.} \bibnamefont{Hossain}},
  \bibinfo{author}{\bibfnamefont{R.}~\bibnamefont{Myrzakulov}},
  \bibinfo{author}{\bibfnamefont{M.}~\bibnamefont{Sami}}, \bibnamefont{and}
  \bibinfo{author}{\bibfnamefont{E.~N.} \bibnamefont{Saridakis}}
  (\bibinfo{year}{2014}), \eprint{1402.6661}.

\bibitem[{\citenamefont{Wetterich}(2004)}]{EDE}
\bibinfo{author}{\bibfnamefont{C.}~\bibnamefont{Wetterich}},
  \bibinfo{journal}{Phys.Lett.} \textbf{\bibinfo{volume}{B594}},
  \bibinfo{pages}{17} (\bibinfo{year}{2004}), \eprint{astro-ph/0403289}.

\bibitem[{\citenamefont{Doran and Robbers}(2006)}]{DR}
\bibinfo{author}{\bibfnamefont{M.}~\bibnamefont{Doran}} \bibnamefont{and}
  \bibinfo{author}{\bibfnamefont{G.}~\bibnamefont{Robbers}},
  \bibinfo{journal}{JCAP} \textbf{\bibinfo{volume}{0606}}, \bibinfo{pages}{026}
  (\bibinfo{year}{2006}), \eprint{astro-ph/0601544}.

\bibitem[{\citenamefont{Wetterich}(1995)}]{CW2}
\bibinfo{author}{\bibfnamefont{C.}~\bibnamefont{Wetterich}},
  \bibinfo{journal}{Astron.Astrophys.} \textbf{\bibinfo{volume}{301}},
  \bibinfo{pages}{321} (\bibinfo{year}{1995}), \eprint{hep-th/9408025}.

\bibitem[{\citenamefont{Birkel and Sarkar}(1997)}]{BS}
\bibinfo{author}{\bibfnamefont{M.}~\bibnamefont{Birkel}} \bibnamefont{and}
  \bibinfo{author}{\bibfnamefont{S.}~\bibnamefont{Sarkar}},
  \bibinfo{journal}{Astropart.Phys.} \textbf{\bibinfo{volume}{6}},
  \bibinfo{pages}{197} (\bibinfo{year}{1997}), \eprint{astro-ph/9605055}.

\bibitem[{\citenamefont{Bean et~al.}(2001)\citenamefont{Bean, Hansen, and
  Melchiorri}}]{BHM}
\bibinfo{author}{\bibfnamefont{R.}~\bibnamefont{Bean}},
  \bibinfo{author}{\bibfnamefont{S.~H.} \bibnamefont{Hansen}},
  \bibnamefont{and}
  \bibinfo{author}{\bibfnamefont{A.}~\bibnamefont{Melchiorri}},
  \bibinfo{journal}{Phys.Rev.} \textbf{\bibinfo{volume}{D64}},
  \bibinfo{pages}{103508} (\bibinfo{year}{2001}), \eprint{astro-ph/0104162}.

\bibitem[{\citenamefont{Doran et~al.}(2007)\citenamefont{Doran, Robbers, and
  Wetterich}}]{A2a}
\bibinfo{author}{\bibfnamefont{M.}~\bibnamefont{Doran}},
  \bibinfo{author}{\bibfnamefont{G.}~\bibnamefont{Robbers}}, \bibnamefont{and}
  \bibinfo{author}{\bibfnamefont{C.}~\bibnamefont{Wetterich}},
  \bibinfo{journal}{Phys.Rev.} \textbf{\bibinfo{volume}{D75}},
  \bibinfo{pages}{023003} (\bibinfo{year}{2007}), \eprint{astro-ph/0609814}.

\bibitem[{\citenamefont{Calabrese et~al.}(2011)\citenamefont{Calabrese,
  de~Putter, Huterer, Linder, and Melchiorri}}]{A2b}
\bibinfo{author}{\bibfnamefont{E.}~\bibnamefont{Calabrese}},
  \bibinfo{author}{\bibfnamefont{R.}~\bibnamefont{de~Putter}},
  \bibinfo{author}{\bibfnamefont{D.}~\bibnamefont{Huterer}},
  \bibinfo{author}{\bibfnamefont{E.~V.} \bibnamefont{Linder}},
  \bibnamefont{and}
  \bibinfo{author}{\bibfnamefont{A.}~\bibnamefont{Melchiorri}},
  \bibinfo{journal}{Phys.Rev.} \textbf{\bibinfo{volume}{D83}},
  \bibinfo{pages}{023011} (\bibinfo{year}{2011}), \eprint{1010.5612}.

\bibitem[{\citenamefont{Reichardt et~al.}(2012)\citenamefont{Reichardt,
  de~Putter, Zahn, and Hou}}]{Re}
\bibinfo{author}{\bibfnamefont{C.~L.} \bibnamefont{Reichardt}},
  \bibinfo{author}{\bibfnamefont{R.}~\bibnamefont{de~Putter}},
  \bibinfo{author}{\bibfnamefont{O.}~\bibnamefont{Zahn}}, \bibnamefont{and}
  \bibinfo{author}{\bibfnamefont{Z.}~\bibnamefont{Hou}},
  \bibinfo{journal}{Astrophys.J.} \textbf{\bibinfo{volume}{749}},
  \bibinfo{pages}{L9} (\bibinfo{year}{2012}), \eprint{1110.5328}.

\bibitem[{\citenamefont{Sievers et~al.}(2013)\citenamefont{Sievers, Hlozek,
  Nolta, Acquaviva, Addison et~al.}}]{Sievers:2013wk}
\bibinfo{author}{\bibfnamefont{J.~L.} \bibnamefont{Sievers}},
  \bibinfo{author}{\bibfnamefont{R.~A.} \bibnamefont{Hlozek}},
  \bibinfo{author}{\bibfnamefont{M.~R.} \bibnamefont{Nolta}},
  \bibinfo{author}{\bibfnamefont{V.}~\bibnamefont{Acquaviva}},
  \bibinfo{author}{\bibfnamefont{G.~E.} \bibnamefont{Addison}},
  \bibnamefont{et~al.} (\bibinfo{year}{2013}), \eprint{1301.0824}.

\bibitem[{\citenamefont{Pettorino et~al.}(2013)\citenamefont{Pettorino,
  Amendola, and Wetterich}}]{A2d}
\bibinfo{author}{\bibfnamefont{V.}~\bibnamefont{Pettorino}},
  \bibinfo{author}{\bibfnamefont{L.}~\bibnamefont{Amendola}}, \bibnamefont{and}
  \bibinfo{author}{\bibfnamefont{C.}~\bibnamefont{Wetterich}}
  (\bibinfo{year}{2013}), \eprint{1301.5279}.

\bibitem[{\citenamefont{Ade et~al.}(2013)}]{PL}
\bibinfo{author}{\bibfnamefont{P.}~\bibnamefont{Ade}} \bibnamefont{et~al.}
  (\bibinfo{collaboration}{Planck Collaboration}) (\bibinfo{year}{2013}),
  \eprint{1303.5076}.

\bibitem[{\citenamefont{Narlikar and Arp}(1993)}]{Na1}
\bibinfo{author}{\bibfnamefont{J.~V.} \bibnamefont{Narlikar}} \bibnamefont{and}
  \bibinfo{author}{\bibfnamefont{H.}~\bibnamefont{Arp}},
  \bibinfo{journal}{Astrophys. J.} \textbf{\bibinfo{volume}{405}},
  \bibinfo{pages}{51} (\bibinfo{year}{1993}).

\bibitem[{\citenamefont{Narlikar}(1977)}]{Na2}
\bibinfo{author}{\bibfnamefont{J.}~\bibnamefont{Narlikar}},
  \bibinfo{journal}{Annals Phys.} \textbf{\bibinfo{volume}{107}},
  \bibinfo{pages}{325} (\bibinfo{year}{1977}).

\bibitem[{\citenamefont{Hoyle and Narlikar}(1966)}]{Na3}
\bibinfo{author}{\bibfnamefont{F.}~\bibnamefont{Hoyle}} \bibnamefont{and}
  \bibinfo{author}{\bibfnamefont{J.~V.} \bibnamefont{Narlikar}},
  \bibinfo{journal}{Proc. R. Soc. London} \textbf{\bibinfo{volume}{A294}},
  \bibinfo{pages}{138} (\bibinfo{year}{1966}).

\bibitem[{\citenamefont{Ayaita et~al.}(2013)\citenamefont{Ayaita, Weber, and
  Wetterich}}]{GNQ10}
\bibinfo{author}{\bibfnamefont{Y.}~\bibnamefont{Ayaita}},
  \bibinfo{author}{\bibfnamefont{M.}~\bibnamefont{Weber}}, \bibnamefont{and}
  \bibinfo{author}{\bibfnamefont{C.}~\bibnamefont{Wetterich}},
  \bibinfo{journal}{Phys.Rev.} \textbf{\bibinfo{volume}{D87}},
  \bibinfo{pages}{043519} (\bibinfo{year}{2013}), \eprint{1211.6589}.

\bibitem[{\citenamefont{Nunes et~al.}(2011)\citenamefont{Nunes, Schrempp, and
  Wetterich}}]{GNQ5A}
\bibinfo{author}{\bibfnamefont{N.~J.} \bibnamefont{Nunes}},
  \bibinfo{author}{\bibfnamefont{L.}~\bibnamefont{Schrempp}}, \bibnamefont{and}
  \bibinfo{author}{\bibfnamefont{C.}~\bibnamefont{Wetterich}},
  \bibinfo{journal}{Phys.Rev.} \textbf{\bibinfo{volume}{D83}},
  \bibinfo{pages}{083523} (\bibinfo{year}{2011}), \eprint{1102.1664}.

\bibitem[{\citenamefont{Mota et~al.}(2008)\citenamefont{Mota, Pettorino,
  Robbers, and Wetterich}}]{GNQ2}
\bibinfo{author}{\bibfnamefont{D.}~\bibnamefont{Mota}},
  \bibinfo{author}{\bibfnamefont{V.}~\bibnamefont{Pettorino}},
  \bibinfo{author}{\bibfnamefont{G.}~\bibnamefont{Robbers}}, \bibnamefont{and}
  \bibinfo{author}{\bibfnamefont{C.}~\bibnamefont{Wetterich}},
  \bibinfo{journal}{Phys.Lett.} \textbf{\bibinfo{volume}{B663}},
  \bibinfo{pages}{160} (\bibinfo{year}{2008}), \eprint{0802.1515}.

\bibitem[{\citenamefont{Baldi et~al.}(2011)\citenamefont{Baldi, Pettorino,
  Amendola, and Wetterich}}]{GNQ8}
\bibinfo{author}{\bibfnamefont{M.}~\bibnamefont{Baldi}},
  \bibinfo{author}{\bibfnamefont{V.}~\bibnamefont{Pettorino}},
  \bibinfo{author}{\bibfnamefont{L.}~\bibnamefont{Amendola}}, \bibnamefont{and}
  \bibinfo{author}{\bibfnamefont{C.}~\bibnamefont{Wetterich}}
  (\bibinfo{year}{2011}), \eprint{1106.2161}.

\bibitem[{\citenamefont{Ayaita et~al.}(2012)\citenamefont{Ayaita, Weber, and
  Wetterich}}]{GNQ9}
\bibinfo{author}{\bibfnamefont{Y.}~\bibnamefont{Ayaita}},
  \bibinfo{author}{\bibfnamefont{M.}~\bibnamefont{Weber}}, \bibnamefont{and}
  \bibinfo{author}{\bibfnamefont{C.}~\bibnamefont{Wetterich}},
  \bibinfo{journal}{Phys.Rev.} \textbf{\bibinfo{volume}{D85}},
  \bibinfo{pages}{123010} (\bibinfo{year}{2012}), \eprint{1112.4762}.

\bibitem[{\citenamefont{Shafi and Wetterich}(1985)}]{HI3}
\bibinfo{author}{\bibfnamefont{Q.}~\bibnamefont{Shafi}} \bibnamefont{and}
  \bibinfo{author}{\bibfnamefont{C.}~\bibnamefont{Wetterich}},
  \bibinfo{journal}{Phys.Lett.} \textbf{\bibinfo{volume}{B152}},
  \bibinfo{pages}{51} (\bibinfo{year}{1985}).

\bibitem[{\citenamefont{Borde et~al.}(2003)\citenamefont{Borde, Guth, and
  Vilenkin}}]{BV}
\bibinfo{author}{\bibfnamefont{A.}~\bibnamefont{Borde}},
  \bibinfo{author}{\bibfnamefont{A.~H.} \bibnamefont{Guth}}, \bibnamefont{and}
  \bibinfo{author}{\bibfnamefont{A.}~\bibnamefont{Vilenkin}},
  \bibinfo{journal}{Phys.Rev.Lett.} \textbf{\bibinfo{volume}{90}},
  \bibinfo{pages}{151301} (\bibinfo{year}{2003}), \eprint{gr-qc/0110012}.

\bibitem[{\citenamefont{Mithani and Vilenkin}(2012)}]{MV}
\bibinfo{author}{\bibfnamefont{A.}~\bibnamefont{Mithani}} \bibnamefont{and}
  \bibinfo{author}{\bibfnamefont{A.}~\bibnamefont{Vilenkin}}
  (\bibinfo{year}{2012}), \eprint{1204.4658}.

\bibitem[{\citenamefont{Copeland et~al.}(1998)\citenamefont{Copeland, Liddle,
  and Wands}}]{QU4}
\bibinfo{author}{\bibfnamefont{E.~J.} \bibnamefont{Copeland}},
  \bibinfo{author}{\bibfnamefont{A.~R.} \bibnamefont{Liddle}},
  \bibnamefont{and} \bibinfo{author}{\bibfnamefont{D.}~\bibnamefont{Wands}},
  \bibinfo{journal}{Phys.Rev.} \textbf{\bibinfo{volume}{D57}},
  \bibinfo{pages}{4686} (\bibinfo{year}{1998}), \eprint{gr-qc/9711068}.

\end{thebibliography}

\end{document}